\documentclass[12pt]{article}
\usepackage{a4wide}
\usepackage{amssymb}
\usepackage{graphicx}
\begin{document}
{\renewcommand{\thefootnote}{\fnsymbol{footnote}}
\begin{center}
{\LARGE  Canonical Relativity\\[2mm] and the Dimensionality of the World\footnote{In: {\em Relativity and the Dimensionality of the
World}, Ed.\ V.~Petkov (Springer, 2007), pp.\ 137--152}}\\
\vspace{1.5em}
Martin Bojowald\\
\vspace{0.5em}
Institute for Gravitation and the Cosmos,\\
The Pennsylvania State
University,\\
104 Davey Lab, University Park, PA 16802, USA\\
\vspace{1.5em}
\end{center}
}

\setcounter{footnote}{0}

\newcommand{\case}[2]{{\textstyle \frac{#1}{#2}}}
\newcommand{\lP}{l_{\mathrm P}}

\newcommand{\md}{{\mathrm{d}}}

\newcommand*{\R}{{\mathbb R}}
\newcommand*{\N}{{\mathbb N}}
\newcommand*{\Z}{{\mathbb Z}}
\newcommand*{\Q}{{\mathbb Q}}
\newcommand*{\C}{{\mathbb C}}

\begin{abstract}
Different aspects of relativity, mainly in a canonical formulation,
relevant for the question ``Is spacetime nothing more than a
mathematical space (which describes the evolution in time of the
ordinary three-dimensional world) or is it a mathematical model of a
real four-dimensional world with time entirely given as the fourth
dimension?'' are presented. The availability as well as clarity of the
arguments depend on which framework is being used, for which
currently special relativity, general relativity and some schemes of
quantum gravity are available. Canonical gravity provides means to
analyze the field equations as well as observable quantities, the
latter even in coordinate independent form. This allows a unique
perspective on the question of dimensionality since the space-time
manifold does not play a prominent role. After re-introducing a
Minkowski background into the formalism, one can see how distinguished
coordinates of special relativity arise, where also the nature of time
is different from that in the general perspective. Just as it is of
advantage to extend special to general relativity, general relativity
itself has to be extended to some theory of quantum gravity. This
suggests that a final answer has to await a thorough formulation and
understanding of a fundamental theory of space-time. Nevertheless, we
argue that current insights into quantum gravity do not change the
picture of the role of time obtained from general relativity.
\end{abstract}

\section{Introduction}

When faced by the question of whether the world is three- or
four-dimensional, the quick answer by a modern-day physicist will most
likely be ``four.'' This is indeed what relativity tells us formally
where space and time are essentially interchangeable: Lorentz
transformations, or their physical manifestations of Lorentz
contraction and time dilatation, show that space and time not only
play similar roles but can even be transformed into each other.  Just
as we can rotate a body in three dimensions to observe all its
extensions, thereby transforming, e.g., its height in width, we can boost
an object\footnote{By ``object'' we will mean a physical system defined by
a set of observable properties such that it can be recognized at
different occurrences in space and in time. Objects will not be
idealized to be point-like or event-like in order to remain unbiased
toward the question of dimensionality. Thus, non-vanishing extensions
of objects in space as well as time are allowed in order to take into
account the necessary unsharpness of measurements needed to verify the
defining properties.} so as to, at least to a certain extent, replace
space-like by time-like extension and vice versa. The qualification
``to a certain extent'' is necessary because even in relativity space
and time are not quite the same but distinguished by the signature of
the space-time metric. By itself, this difference in signature is not
sufficient reason to deny time the same ontological status as space.

There are, however, differences between the usual treatment of space
and time in physics going beyond relativity, although they are usually
presupposed in relativistic discussions. In order to answer the
question of the dimensionality of the world from the viewpoint of
relativity, such hidden assumptions have to be uncovered and analyzed,
or avoided altogether. Some of these issues lie at the forefront of
current physics and still await explanation. For instance, while we
can, and have to, limit objects to finite spatial extensions, we have
no means to limit their time extensions safe for transformations such
as particle decay or other reactions. Even though objects may change
in time, they never cease to exist completely. There seems to be a
simple reason for that: conservation laws. We simply cannot limit an
object's extension in time because, e.g., its energy must be
conserved. Thus, the object could be transformed into something else
of the same energy but not removed completely. Such laws are derived
as consequences of symmetries which first give local conservation laws
in terms of currents. Going from local to global conservation laws, as
they are required for an explanation of the persistence of objects in
time, is a further step and requires additional assumptions. As the
following more detailed discussion shows, conservation laws cannot be
used to explain the difference of spatial and time-like extensions,
for the derivation itself distinguishes space from time.

One starts with a local equation such as\footnote{We follow the
abstract index notation common in general relativity (see, e.g.,
\cite{Wald}). Indices $a,b,\ldots$ refer to space-time while indices
$i,j,\ldots$ used later refer only to space. Repeated indices
occurring once raised and once lowered are summed over the
corresponding range $0,1,2,3$ for space-time indices and $1,2,3$ for
space indices. The covariant derivative compatible with a given
space-time metric $g_{ab}$ is denoted by $\nabla_a$ which for
Minkowski space reduces to the partial derivatives $\partial_a$ in
Cartesian coordinates.}  $\nabla^aT_{ab}=0$ for the energy-momentum
tensor $T_{ab}$. If the space-time metric is sufficiently symmetric
and allows a Killing vector field $\xi^a$ satisfying ${\cal
L}_{\xi}g_{ab}=\nabla_a\xi_b+\nabla_b\xi_a=0$, the current
$j_a=T_{ab}\xi^b$ is conserved: $\nabla^aj_a=0$. At this stage, the
only difference between space and time enters through the signature of
the metric. A global conservation law is then derived by integrating
the local conservation equation over a space-time region bounded by
two spatial surfaces $\Sigma_1$ and $\Sigma_2$ and some boundary $B$
which could be at infinity. Stokes theorem then shows that the
quantity $\int j_a\md S^a$ is the same on $\Sigma_1$ and $\Sigma_2$
and thus conserved, {\em provided that all fields fall off
sufficiently rapidly toward the boundary $B$}. Thus, one already has
to assume physical objects to be of finite spatial extent before
obtaining a global conservation law, while there is no such
restriction for the time-like extension.\footnote{In this discussion
we understood, as usually, that space-time is Minkowski as in special
relativity. The energy conservation argument in our context works
better if one considers instead a universe model with compact spatial
slices, such as an isotropic model with positive spatial curvature, or
a compactification of isotropic models of non-positive curvature. This
requires one to go beyond special relativity, as we will do later on
for other reasons, and to allow non-zero curvature or non-trivial
topology. From our perspective, closed universe models are
conceptually preferred because space is already finite without
boundary such that non-spacelike boundaries are not needed to derive
global conservation laws from local ones.} If fields do not vanish at
$B$, one interprets $\int_B j_a \md S^a$ as the flux into or out of
the spatial region evolving from $\Sigma_1$ to $\Sigma_2$. However,
this different interpretation of $\int j_a\md S^a$ as conserved
quantity on $\Sigma_1$ and $\Sigma_2$ and as flux on $B$ treats space
and time differently, corresponding to the non-relativistic
decomposition of energy-momentum in energy and momentum. This
different treatment is not implied by the theory but put in by
interpreting its objects. The issue of a limited spatial extent versus
unconstrainable duration of objects thus remains and has to be faced
even before coming to conservation laws.

\begin{figure}
\begin{center}
\includegraphics[height=4cm,keepaspectratio]{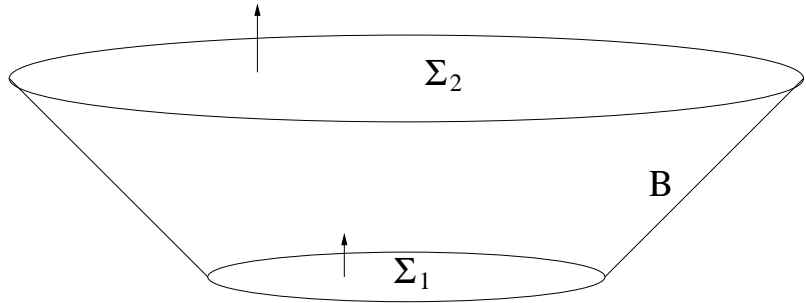}
\caption{Causal diagram of space-time region integrated over to derive
global conservation laws.}
\end{center}
\end{figure}

For this reason, it seems to be potentially misleading to consider
objects in space-time such as point particles or their worldlines to
address the dimensionality of the world, for there are already
presuppositions about space and time involved. Indeed, from this
perspective a worldline, or the world-region of an extended object,
seems inadequate for a relativistic treatment. It would be more
appropriate to use only either space-time events or bounded
four-dimensional world-regions of extended objects. This already
indicates a possible answer to the question of dimensionality:
events
are zero-dimensional and to be considered as idealizations just as
their analogs of point particles. The only option then is to consider
bounded world-regions\footnote{Indeed, to analyze an object by
whatever means we not only need to capture it at one time --- which is
virtually impossible, anyway --- but also hold and observe it for some
time. Observations thus always refer to some finite extension in time
during which we must be able to recognize the system. A good example
can be found in particle physics where too short decay times imply
that particles appear rather as resonances without sharp values for
all their properties. This is a consequence of uncertainty and thus
quantum theory which we will come back to later. Even though common
terminology often assigns the object status to an isolated system at a
given time, evolving and possibly changing, observations always
consider world-regions which could be assigned the object status as
well. This non-traditional use of the term ``object'' is probably
discouraged because it is too observer-dependent: it is the observer
who decides when to end the experiment and thus determines the
time-like extension of the space-time region. However, while the
classical world allows us to draw sharp spatial boundaries and thus
seems to imply individualized spatial objects, this is no longer
possible in a quantum theory.  Drawing the line around spatial objects
is then a matter of choice, too, comparable to limiting the duration
of an observation.} as physical objects, which are four-dimensional.

It is difficult to follow these lines toward a clear-cut argument for
the four-dimensionality of the world due to our limited understanding
of the nature of time. An alternative approach is to disregard objects
in space-time and rather consider the relativistic physics of
space-time itself. For this, we need general relativity which,
compared to special relativity, has the added advantage of removing
the background structure given by assuming Minkowski space-time. As
backgrounds can be misleading, if possible one should consider the
more general situation and then see how special situations can be
re-obtained.

The following sections collect possible ingredients which can be
helpful in the context of dimensionality. There are different
formulations of general relativity, covariant and canonical ones,
which apparently reflect the possible interpretations of a
four-dimensional versus a three-dimensional world: While covariant
field equations are given on a space-time manifold, the canonical
formulation starts with a slicing of space-time in a family of spatial
slices. Canonical fields are then defined on space and evolve in time,
suggesting a three-dimensional world with an external time parameter
\cite{ADM}. Nonetheless, the question of dimensionality cannot be
answered easily because, for one thing, the formulations are
mathematically equivalent. In what follows, we will mainly use
canonical relativity so as to see if it indeed points to a
three-dimensional world or, as the covariant formulation, a
four-dimensional one.  Covariant formulations are also best suited to
understand the relativistic kinematics and dynamics. After this
general exposition we will specialize the formalism to Minkowski space
in order to see which freedom is eliminated in special relativity
compared to general relativity and how this can change the picture of
time. We end with a brief discussion of dynamical consequences of
general relativity as well as some comments on quantum aspects.

\section{Canonical Relativity}

The signature of the metric also has implications for the form of
relativistic field equations on a given space-time which are
hyperbolic rather than elliptic. This means that a reasonable set-up
for solving these equations is by an initial value problem: for given
initial values on space at an initial time one obtains a unique
solution.  For our purposes, this aspect is not decisive because we
could interpret initial values as corresponding to objects placed in
space before starting an observation, thus corresponding to a
three-dimensional world, or simply as labels to distinguish solutions
which themselves play the role of objects of a four-dimensional
world.\footnote{In fact, this is clearly brought forward by the
canonical formulation where one can either specify states by a phase
space given by initial data on the initial surface, or by the
so-called covariant phase space consisting of entire solutions to the
field equations. In both cases, the phase space is endowed with a
symplectic structure, and the formulations are equivalent.} The choice
is then just a matter of convenience.

\subsection{ADM Formulation}

For the field equations of the metric itself the situation is more
complicated and crucially different (see \cite{CauchyProblem} for the
general relativistic initial value problem). Einstein's equations
correspond to ten field equations for the ten components of the
space-time metric $g_{ab}$, a symmetric tensor. However, there are
only six evolution equations containing time derivatives only of some
components while the remaining equations are elliptic and do not
contain time derivatives. Although there is no fixed coordinate
system, it is meaningful to distinguish between time and space
derivatives because, due to the signature of the metric, they are
related to vector fields of negative and positive norm squared,
respectively. Time evolution is described by an arbitrary timelike
vector field $t^a$ while spatial slices of space-time are introduced
as level surfaces $\Sigma_t\colon t={\rm const}$ of a time
function $t$ such that $t^a\partial_at=1$. The space-time metric
$g_{ab}$ induces a spatial metric $h_{ab}(t)$ on each slice $\Sigma_t$
as well as covariant spatial derivatives. The spatial metric is most
easily expressed as
\begin{equation}\label{metric}
 h_{ab}=g_{ab}+n_an_b
\end{equation}
where $n_a$ is the unit future-pointing timelike co-normal to a
slice. These are only six independent components because $h_{ab}$ is
degenerate from the space-time point of view: $n^ah_{ab}=0$. These are
also precisely the components of the space-time metric whose time
derivatives\footnote{Time derivatives are understood as Lie
derivatives with respect to the time evolution vector field $t^a$.}
appear in Einstein's equations. 

\begin{figure}
\begin{center}
\includegraphics[height=5cm,keepaspectratio]{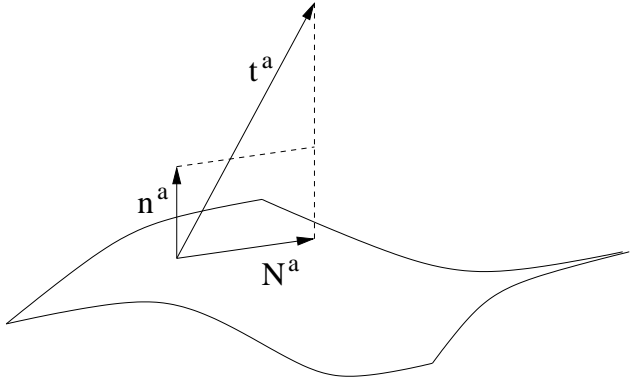}
\caption{Decomposition of the time evolution vector field $t^a$ into
the shift vector $N^a$ and a normal contribution $Nn^a$.}
\end{center}
\end{figure}

At this point, we may view the equations as describing the evolution
of a three-dimensional quantity $h_{ab}$ in an external time parameter
$t$. The remaining four space-time metric components encode the
freedom in choosing the time evolution vector field, which can be
parameterized as $t^a=Nn^a+N^a$ with components usually called lapse
function $N$ and shift vector $N^a$ such that $n_aN^a=0$. They are
indeed metric components since (\ref{metric}) implies
\begin{eqnarray}
 g^{ab} &=&-n^an^b+h^{ab}=
-\frac{1}{N^2}(t^a-N^a)(t^b-N^b)+h^{ab}\nonumber\\
 &=& -\frac{1}{N^2}t^at^b+\frac{1}{N^2} (t^aN^b+N^at^b)+
h^{ab}-\frac{1}{N^2}N^aN^b \label{invmetric}.
\end{eqnarray}
The time-time component of the inverse space-time metric is thus
$-N^{-2}$ while time-space components are $N^{-2}N^a$.  These
components enter the field equations, too, but they are not dynamical
in the sense that they would have evolution equations determining
their time derivatives. In addition to the six evolution equations for
$h_{ab}$,
\begin{equation}
 \dot{\pi}^{ab}=f[h_{ab},\pi^{ab},N,N^a]
\end{equation}
for the momenta $\pi^{ab}[\dot{h}_{cd},N,N^c]$ conjugate to $h_{ab}$,
there are then four constraint equations
\begin{equation}
 C[h_{ab},\pi^{ab}]=0 \quad\mbox{and}\quad C^a[h_{bc},\pi^{bc}]=0
\end{equation}
which are of elliptic nature and restrict the values of the dynamical
fields at any spatial slice (independently of $N$ and $N^a$).

A possible interpretation is that there are six fields $h_{ab}$ on
space which change in time as governed by the evolution equations,
depending on four prescribed but arbitrary auxiliary fields $N$ and
$N^a$. This would be a mixed viewpoint as far as dimensionality is
concerned because $h_{ab}$ would look like spatial objects while $N$
and $N^a$ would have to be {\em prescribed} as functions of space and
time but are not evolving in time. The system is thus rather
four-dimensional since $N$ and $N^a$ have to be functions on a
four-dimensional space and determine the evolution of $h_{ab}$ for
which only initial values on space are needed. One can save the
three-dimensional interpretation by considering $N$ and $N^a$ as
external functions for the evolution system of $h_{ab}$, but this has
the drawback that there would be no predictivity and no
uniqueness of solutions in terms of initial values for the dynamical
fields, as solutions also depend on choices of $N$ and $N^a$.

Here, the completely four-dimensional view is much more attractive: We
not only have to choose the functions $N$ and $N^a$ but can also
supplement the dynamical equations for $h_{ab}$ by evolution equations
for space-time {\em coordinates}. This is indeed possible, for if we
choose four functions $N$ and $N^a$ and ask that they play the
role of space-time metric components as they enter the canonical
equations (\ref{invmetric}) we have the transformation laws
\begin{eqnarray}
 -\frac{1}{N(t,x^i)^2} &=& q^{bc}(x')\partial'_bt\partial'_ct\\
 N(t,x^i)^{-2}N^j(t,x^i)&=&q^{cd}(x')\partial'_ct\partial'_dx^j
\end{eqnarray}
from an arbitrary (inverse) space-time metric $q^{ab}$ to the new
functions. These transformations can be interpreted as evolution
equations for the space-time coordinates which are then fixed by the
choice of $N$ and $N^a$ in terms of coordinates on an initial spatial
surface.\footnote{Solutions $x(x')$ depend on the auxiliary metric
$q^{ab}$ as well, as they should since no change in coordinates is
necessary if $q^{ab}$ had already been of canonical form with the
chosen $N$ and $N^a$.} With this interpretation, we obtain, for given
initial values, unique solutions to our evolution equations up to
changes of coordinates, corresponding to a change in the free
functions $N$ and $N^a$.

The functions $N$ and $N^a$ which must be defined on a
four-dimensional manifold thus determine coordinates $x^a$ such that
canonical field equations for $h_{ab}$ result. In general, there is no
way to split this globally into a time coordinate $t$ and
space-coordinates $x^i$ which one would need for a three-dimensional
evolution picture of $h_{ab}$. Thus, a four-dimensional interpretation
results. This attractive viewpoint is available only if we take as
physical object on which field equations are imposed the entire
space-time and not just the metric on spatial slices. We clearly have
to consider general relativity which gives field equations for the
metric and allows us to perform arbitrary coordinate changes, not just
Lorentz transformations. Here, getting rid of the Minkowski background
of special relativity, corresponding to a synchronization of rigid
clocks and rulers throughout space-time under the assumption of the
absence of a gravitational field, is required.

\subsection{Relational Observables}

Since we obtain a unique solution to the Einstein equations only up to
arbitrary changes of space-time coordinates, predictivity requires
observable quantities to be coordinate independent, too. While
coordinates are usually used in explicit calculations, values of
observable quantities must not change when transforming to different
coordinates. Abstractly, one can also formulate the concept of an
observable in an explicitly coordinate-free manner leading to
relational observables: evolution is then measured not with respect to
coordinate time but with respect to other geometrical or matter
quantities. While this is appealing conceptually, it can be hard to do
explicitly.\footnote{``Such a question can, we are assured, always be
answered from a sufficient set of initial data, though the performance
of this task may call for considerable mathematical agility.''
\cite{BergmannTime}.} For instance, in a cosmological situation one can measure
how the value of a matter field changes with respect to a change in
the total spatial scale or volume. Since in such a picture coordinates
are eliminated, an alternative view on the question of dimensionality
is possible. It also allows us to show, as we will see, how Minkowski
space is recovered and what is special about special relativity.

Coordinate changes on a manifold imply transformations for fields such
as $g_{ab}$ on that manifold. Observable quantities then must be
expressions formed by the fields of a theory being invariant under any
change of coordinates. Simple examples are integrals of
densities\footnote{A density is a mathematical object transforming in
the same manner as $\sqrt{\det g}$ under changes of coordinates such
that its coordinate integration is well-defined.} over the whole
space-time manifold, but they are too special and do not give one
access to local properties. A more general, abstract way of
constructing observables is as follows:\footnote{This idea goes back
to Komar and Bergmann \cite{KomarObs,BergmannTime} and has more
recently been elaborated and used in, e.g.,
\cite{GeomObs1,GeomObs2,MiniQuant,PartialCompleteObs,PartialCompleteObsII,DittrichThesis,GPSCoord,EffectiveObs}.}
We use the group of transformations of our basic fields corresponding
to coordinate changes $x^a\mapsto x'{}^a(x^b)$, which in our case is
the group of space-time diffeomorphisms.\footnote{Space-time
diffeomorphisms are in general not in one-to-one correspondence with
coordinate changes. For our purposes, local considerations are
sufficient where this identification can be made. A local coordinate
change is then infinitesimally given by $x^a\mapsto x^a+\xi^a(x^b)$
where the vector field $\xi^a$ is of compact support, and the same
vector field generates a diffeomorphism.}  In an explicit realization,
group elements would have infinitely many labels corresponding to four
functions on space-time, or a space-time vector field $\xi^a(x)$. A
relational observable requires one to choose a quantity $f$ to be
measured with respect to as many other functionals $\Phi^a_x$ of the
basic fields as there are parameters of the group.\footnote{For a
specific example, $f$ could be a matter field and $\Phi^a_x$ the
spatial volume $\det h_{ab}$ in an isotropic cosmological model. Here,
the infinite number of variables $\Phi^a_x$ is reduced to only one by
the high degree of symmetry. This corresponds to the fact that only
spatially constant time reparameterizations respect the
symmetry. Thus, the label $a$ disappears because spatial coordinates
cannot be changed in a relevant manner (they can be rescaled in some
cases, without affecting the basic fields), and $x$ disappears due to
spatially constant reparameterizations. We will come back to possible
reductions in the number of independent variables from the
counterintuitive infinite size in the following subsection.}  These
functionals $\Phi^a_x$ will be called internal
variables,\footnote{Often, ``internal time'' or ``clock variables''
are used in this context, as these quantities are commonly employed to
discuss the problem of time. However, since they do not only refer to
time and it is even unclear in which sense time is involved, we prefer
a neutral term.}  for gravity labeled by the space-time index $a$ and
a point $x$ in space-time. This corresponds to the freedom in labels
of the diffeomorphism group. From $f$ and $\Phi^a_x$ we construct an
observable\footnote{The notation, similar to that in
\cite{DittrichThesis,PartialCompleteObs}, is quite loaded and
indicates that $F[\Phi^a_x]$ is a relational object telling us how $f$
changes under changes of the internal variables $\Phi^a_x$. The answer
depends parametrically on infinitely many real numbers $\phi^a_x$: for
each fixed set of these parameters, $F[\Phi^a_x]_{\phi^a_x}$ gives
coordinate independent information on the relational behavior as a
functional of the basic field $g_{ab}$.} $F[\Phi^a_x]_{\phi^a_x}$ as a
functional of the basic fields parameterized by time values $\phi^a_x$
as real numbers: to compute the evaluation
$F[\Phi^a_x]_{\phi^a_x}(g_{ab})$ of the observable on a given set of
basic fields $g_{ab}$ we first find a coordinate transformation for
which $g'_{ab}$ becomes such that $\Phi^a_x(g'_{ab})=\phi^a_x$ equal
the chosen time parameters. The value of the observable is then
defined to be the original function $f(g'_{ab})$ evaluated in this
transformed set of basic fields. For any set of time variables
$\phi^a_x$ one obtains a functional of the basic fields $g_{ab}$. This
clearly results in an observable independent of the system of
coordinates, and is well-defined at least for certain ranges of the
fields and parameters involved.\footnote{Global issues, as always in
general considerations for general relativity, are much more difficult
to handle.}

The observable, interpreted as measuring the change of $f$
relationally with respect to the internal variables rather than with
respect to coordinates does, however, depend on the parameters $\phi^a_x$
which crucially enter the construction. One is rather dealing with a
family of observables labeled by these parameters. While one obtains
an observable for each fixed set of parameters, its interpretation
would be complicated and loose any dynamical information of
change. This is probably one of the clearest indications for the
dimensionality of the world from a mathematical point of view: What we
are constructing directly are relational observables depending on
parameters $\phi^a_x$, roughly corresponding to a set of
worldlines. While this can be restricted to fixed
parameters,\footnote{In fact, even though the $\phi^a_x$ are sometimes
called ``time parameters,'' only for one value of $a$ does it really
correspond to an infinity of times while the remaining parameters are
space parameters. Having also space-parameters is actually an
advantage in light of our earlier discussion where entire
world-regions rather than any kind of lower-dimensional object were
preferred. Such a world-region is then spanned by suitable ranges of
all the parameters $\phi^a_x$.} it would be a secondary step. Moreover,
if all parameters are fixed, also spatial dependence is eliminated; in
such a case we end up only with non-local observables. The primary
observable quantities are thus not spatial at all but rather give, in
an intricate, relational manner, a four-dimensional world.

On second thought, there seems to be a problem because we have
infinitely many parameters. From special relativity, or any kind
of non-relativistic physics, we would expect only one time parameter
in addition to three space parameters as independent
variables.\footnote{This refers strictly only to one observer. In
special relativity one considers time and space coordinates between
boosted observers. For a given observer, the synchronization
conditions of special relativity imply that there is only one time and
three space parameters.}  On the other hand, special relativity is
obtained from general relativity by introducing a background given by
Minkowski space-time. Physically, this corresponds to synchronizing
all clocks to measure time (and using a fixed set of rulers to measure
lengths). When all clocks are synchronized, there is only one time
parameter, and so it is not surprising after all that general
relativity, lacking a synchronization procedure, requires infinitely
many parameters $\phi^a_x$ for its observable quantities. The
mathematical situation is thus in agreement with our physical
expectations. We will now make this more explicit by showing how a
Minkowski background can be re-introduced.

\subsection{Recovering the Minkowski Background}

The synchronization procedure can be implemented directly for general
relational observables, clearly showing the reduction from infinitely
many parameters to only one time coordinate. This brings us to the
promised recovery of special relativity by re-introducing the
Minkowski background and illustrates the relation between the
infinitely many parameters of relational observables and the finite
number of coordinates in Minkowski space. We make use of expressions
derived recently for general relativity
\cite{PartialCompleteObs,PartialCompleteObsII,DittrichThesis}. We
assume that four internal field variables $\Phi^a(x)$ have been
chosen, having conjugate momenta $\Pi_a$ in a canonical formulation,
which in a space-time region we are interested in are monotonic
functions of $x^b$. For simplicity, we assume that these variables are
four scalar fields which are already present in the theory, rather
than more complicated functionals of basic fields such as curvature
scalars used in \cite{KomarObs,BergmannTime}. Moreover, we ignore
their dynamics, i.e.\ assume that there are no potentials, since our
aim here is to reconstruct the non-dynamical Minkowski
space-time. Geometrically, the momenta are given by the (density
weighted) derivatives of the internal variables along the unit normal
to spatial slices,
\begin{equation}\label{momentum}
 \Pi^a=\sqrt{\det h} n^b\partial_b \Phi^a\,.
\end{equation}
This determines the rate by which the fields change from slice
to slice.  In a region of monotonic fields, we can thus view
$x^a\mapsto \Phi^a(x)$ as a coordinate transformation and transform our
metric accordingly, observing (\ref{metric}) and (\ref{momentum}):
\begin{equation}
 g'{}^{ab} = \partial_c\Phi^a\partial_d\Phi^b(h^{cd}-n^cn^d)= 
 \partial_c\Phi^a\partial_d\Phi^bh^{cd}-\Pi^a\Pi^b/\det h
\end{equation}
or, splitting into time and space components,
\begin{eqnarray}
 g'{}^{00} &=& \partial_i\Phi^0\partial_j\Phi^0 h^{ij}
 -\Pi^0\Pi^0/\det h \label{transa}\\
 g'{}^{i0} &=& \partial_j\Phi^0\partial_k\Phi^ih^{jk}- 
\Pi^i\Pi^0/\det h\label{transb}\\
 g'{}^{ij} &=& \partial_k\Phi^i\partial_l\Phi^j h^{kl}- 
 \Pi^i\Pi^j/\det h\,.\label{transc}
\end{eqnarray}

First, we suppress the components $\Phi^i$ to bring out the role of
time which will be played by $\Phi^0$. We thus assume that the spatial
metric $h^{ij}$ is already given by $\delta^{ij}$ as in Minkowski
space in its standard coordinate representation. Under the remaining
transformation corresponding to $\Phi^0$, the original spatial metric
$h^{ij}$ is transformed to the new spatial metric $g'{}^{ij}$ which to
preserve Minkowski space should also equal $\delta^{ij}$. Since we
suppressed the spatial parameters $\Phi^i$, we need to require
$\Pi^i=0$ such that the spatial coordinate system is fixed in
time.\footnote{Thus, our set of rulers does not change in time.} Time
synchronization then implies that $\Phi^0$ does not depend on spatial
coordinates, so also $g'{}^{i0}=0$ is of Minkowski form. For the final
component of the metric $g'{}^{ab}$ we obtain $g'{}^{00}=-(\Pi^0)^2$
which is of Minkowski form for $\Pi^0=1$.  These conditions can be
summarized by saying that spatial coordinates do not change in time
($\Pi^i=0$), and time progresses at the same constant pace everywhere
($\Pi^0=1$).

For instance from the construction of relational observables in
\cite{PartialCompleteObs} it follows that with such a choice of clock
variables a relational observable takes the form
\begin{equation} \label{ObsFixed}
 F[\Phi^a_x]_{\phi^0}=\sum_{k=0}^{\infty}\frac{1}{k!} \dot{f}(\Phi^i)
(\phi^0-\Phi^0)^k= f(\Phi^i,\phi^0-\Phi^0)
\end{equation}
where the dot refers to the change in $f$ under a change of the time
field $\Phi^0$. If we identify space-time coordinates with $\Phi^a$, any
function will be observable since the background is completely fixed
for a given observer.

This refers to one observer who has performed a full
synchronization. If we change the observer, we obtain the usual
Lorentz transformations and between two different observers time does
certainly not proceed at the same pace. To see this, we now allow all
four functions $\Phi^a$ to be non-trivial. We want to describe a
situation where one synchronized observer is given by a system of the
form just derived, such that $h^{ij}=\delta^{ij}$, whose internal
variables we now call $\Psi^a$. From there, we transform to a new
system of internal variables $\Phi^a(\Psi^b)$ such that also the
metric $g'{}^{ab}$ is Minkowski for the new synchronized
observer. Thus, the right hand sides of (\ref{transa}),
(\ref{transb}), (\ref{transc}) must be $\Psi^a$-independent and only
linear functions $\Phi^a$ are allowed:
\begin{eqnarray}
 \Phi^i &=& \omega^i_j \Psi^j+\alpha^i \Psi^0\\
 \Phi^0 &=& \beta_i\Psi^i+\gamma \Psi^0\,.
\end{eqnarray}
Derivatives in (\ref{transa}), (\ref{transb}), (\ref{transc}) are now
taken with respect to $\psi^i$, and
$\Pi^a=\partial\Phi^a/\partial\Psi^0$.  From (\ref{transa}) we then
obtain $g'{}^{00}=-1=\beta_i\beta_j
\delta^{ij}-\gamma^2$ such that 
\begin{equation} \label{gamma}
 \gamma=\sqrt{1+|\beta|^2}
\end{equation}
where $|\cdot|$ denotes the norm of vectors in $h_{ij}=\delta_{ij}$. From
Eq.~(\ref{transb}) in the form
\[
  g'{}^{i0} = \partial_j\Phi^0\partial_k\Phi^i\delta^{jk}- \Pi^i\Pi^0
\]
we have $0=\beta_j\omega^i_k\delta^{jk}-\alpha^i\gamma$ such that
\begin{equation} \label{alpha}
 \alpha^i=\frac{\beta_j\omega^i_k \delta^{jk}}{\gamma}\,.
\end{equation}
Finally, Eq.~(\ref{transc}) in the form
\[
 g'{}^{ij} = \partial_k\Phi^i\partial_l\Phi^j\delta^{kl}-\Pi^i\Pi^j
\]
implies
\begin{equation}\label{rot}
 \delta^{ij}= \omega^i_k\omega^j_l\delta^{kl}-\alpha^i\alpha^j\,.
\end{equation}
Defining 
\begin{equation}
 \rho^i_j:=\omega^i_j-\frac{\alpha^i\beta_j}{1+\gamma}\,,
\end{equation}
for which we have
\begin{eqnarray*}
 \rho^i_k\rho^j_l\delta^{kl} &=& \omega^i_k\omega^j_l\delta^{kl}-
\frac{\omega^i_k\alpha^j\beta^k}{1+\gamma}-
\frac{\omega^j_k\alpha^i\beta^k}{1+\gamma}+
\frac{\alpha^i\beta_k\alpha^j\beta^k}{(1+\gamma)^2}\\
 &=& \omega^i_k\omega^j_l\delta^{kl}- \alpha^i\alpha^j
\end{eqnarray*}
using (\ref{gamma}) and (\ref{alpha}), shows that the freedom in
$\omega^i_j$ is given by an orthogonal matrix $\rho^i_j$.  Thus, only
a vector $\beta^i$ and a rotation $\rho^i_j$ can be chosen freely to
specify a transformation.  The remaining coefficients $\alpha^i$ and
$\gamma$ are then fixed by (\ref{alpha}) and (\ref{gamma}). This is
easily recognized as the usual coefficients of Lorentz transformations
if we only identify $\beta^i=v^i/\sqrt{1-v^2/c^2}$ and use
$\rho^i_j$ as the rotational part of the transformation.

Allowing different synchronized observers, observable functions as in
(\ref{ObsFixed}) have to be Lorentz invariant and are not arbitrary.
Completely arbitrary, non-synchronized observers then require the
general relativistic situation with complicated relational expressions
for observables.

From our perspective, this shows that the usual space and time
parameters one has in special relativity are what is left after fixing
all but finitely many of the infinitely many parameters $\phi^a_x$. These
infinitely many parameters occur automatically when one attempts to
write observables in a relational manner. In general, none of these
parameters is distinguished as a possible time parameter to describe
the evolution of a three-dimensional world. In the relational picture,
thus, only the four-dimensional option is available.

\section{Challenges and Resolutions}

The canonical structure of relativity and an analysis of what is
observable thus gives good reasons for the four-dimensionality of the
world. Some difficulties certainly remain because, for one thing, we
considered only local regions and had to assume that we can find
functions $\Phi^a_x$ which are monotonic there. In order to describe the
whole space-time in this manner we would need globally monotonic
functions which may be difficult to find in general. For strictly
physical purposes such a global description is also an
over-idealization because all observations we can ever make are
restricted to some bounded region of space-time, however big this
region may be in cosmological observations.  There are more severe
potential challenges to this picture, one resulting from properties of
general relativity not considered so far, and the other resulting from
quantum theory.

\subsection{Singularities}

Locally, solutions to Einstein's field equations always exist and
determine the space-time metric as well as manifold. This played a
crucial role in our arguments given so far where we wanted to
eliminate backgrounds and consider dynamical space-time. These
equations are, however, non-linear and so global aspects are more
difficult to control. One consequence is that most solutions which we
think are relevant for what we observe are singular when extrapolated
in general relativity. They allow one to describe space-time only for
a finite amount of proper time for some, and in some cases all,
observers after which the classical theory breaks down
\cite{SingTheo}. This is usually accompanied by a divergence of
curvature, but in any case represents a finite boundary to space-time.

If the theory does not allow us, even in principle, to extend
solutions arbitrarily far in one direction, it may be difficult to
view this direction as a dimension of the world. Here, the
three-dimensional viewpoint seems more suitable because we would
simply have to deal with space and objects in time, described by the
theory for some finite range of time. To be sure, there are also
solutions where space is finite, but even if there are such boundaries
space-time can usually be extended and they are thus
artificial.\footnote{There can also be boundaries to space arising
from singularities where space-time cannot be extended in spatial
directions. Such time-like singularities, however, do not generically
arise in relevant cosmological or black hole solutions and thus can be
ignored here. In homogeneous cosmological models, from which most of
the cosmological intuition is derived, such time-like singularities
are ruled out by the assumption of homogeneity (be it a precise or
approximate symmetry) while for black holes time-like singularities
arise for negative mass where the singular behavior is even welcome to
rule out negative mass and argue for the stability of Minkowski
space. Other black hole solutions where time-like singularities arise,
such as the Reissner--Nordstrom solution for electrically charged
black holes in vacuum, are unstable to the addition of matter. Generic
singularities are then space-like or null
\cite{SpacelikeSing,NullSing}.}  This is not the case with
singularities. If we are interested in a four-dimensional
interpretation, then, we will have to deal with fundamental
limitations to the extension of four-dimensional objects, including
space-time itself.

\subsection{Quantum Aspects}

Just as it was helpful to embed special relativity into general
relativity for a wider viewpoint, the classical description is itself
incomplete not the least because it leads to space-time
singularities. This requires a corresponding extension of general
relativity to a quantum theory of gravity. But even before this stage
is reached, quantum properties do have a bearing on some of the
arguments that can be used to decide on the dimensionality of the
world. For instance, the four-dimensional interpretation is
advantageous because it embodies the fact that we have to recognize an
object in order to denote it as such, showing that the time extension
plays a central role in assigning object status. Such a recognition is
not possible in quantum mechanics where identical particles are
indistinguishable. We can then never be sure that a particle we
recognize is the same one we saw before, and so assigning object
status to worldlines or world-regions would not make sense unless all
identical particles are subsumed in one and only one object.

\subsection{Resolutions}

These puzzles are resolved easily if one just considers suitable
combinations of quantum theory with special and general relativity,
respectively. Combining quantum theory with special relativity leads
to quantum field theory where indeed the particle concept is weakened
compared to the classical or quantum mechanical picture. There is not
a collection of individual but indistinguishable particles, but a
field whose excitations may in some cases be interpreted as
particles. Thus indeed, one is treating all identical particles as one
single object, the corresponding field, and any problem with
recognizability is removed automatically. The field is a function on
space-time or a world-region, a four-dimensional object.\footnote{In
algebraic quantum field theory one considers algebras associated with
world-regions of ``diamond'' shape as the basic objects, so also here
it is bounded regions in space-time determining what objects in the
theory are.}

Singularities of general relativity pose a more complicated problem,
but there are indications that they, too, are automatically dealt with
when the underlying classical theory, this time general relativity, is
combined with quantum theory. While the classical space-time picture
breaks down at a singularity, several recent investigations have shown
that quantum geometry continues to be well-defined, albeit in a
discrete manner
\cite{Sing,HomCosmo,Spin,BHInt,ModestoConn,SphSymmSing}. One can then
extend the classical space-time through a quantum region, or view
space-time as fundamentally described by a quantum theory of gravity
which reduces to general relativity in certain limits when curvature
is not too large.

Indeed, background independent versions of quantum gravity are not
formulated on a space-time manifold such that the question of whether
the three- or four-dimensional viewpoint should be taken does not
really arise at all. One is either dealing with space-time objects
directly, such as in discrete path integral approaches, or employs a
canonical quantization where the central object is a wave function on
the space of geometries and observables are relational as discussed
before. In quantum gravity, the four-dimensional, relational viewpoint
is thus even more natural than in classical gravity. It is also
crucial for the results on non-singular behavior which are based on
the relational behavior of wave functions or other quantities which
now play the role of basic objects. Extensions beyond classical
singularities can then be provided by considering the range of
suitable internal variables and their quantizations: The relational
dependence can, and in all cases studied so far will, continue through
stages where one would classically encounter a singularity. This is
much more robust than looking at possible modifications of field
equations and corresponding extensions of space-times in coordinate
form which have turned out to be non-generic if available at all.

\section{Conclusions}

The question of whether a theoretical object is just a mathematical
construct or a real physical thing is always difficult to address in
physics. Often, the answer depends on what theory is used,
which itself depends on current available knowledge. Not just the
theoretical structure needs to be understood well but also its
ontological underpinning. This is notoriously difficult if space and
time are involved, and often hidden assumptions already enter
constructions.

In such a situation, it is best to make use of as flexible a framework
as possible and to eliminate any background structure. Thus, we
focused on general relativistic dynamics rather than special
relativistic kinematics. We have highlighted some relevant
consequences using a canonical formulation. Canonical formulations are
often perceived as not being preferable because they break manifest
covariance. However, they also offer well structured mathematical
formulations and can be particularly illuminating for the dynamical
behavior.

In particular, canonical techniques allow, or even require one to discuss
observables in a coordinate free manner. This leads to a relational
description where no coordinates are used but instead field values are
related to values of other fields to retrieve observable
information. Usually, these quantities take the form of families of
functionals parameterized by real numbers (most generally, infinitely
many ones). In contrast to coordinates, these parameters do not
distinguish between space and time and even the signature of a
space-time metric is irrelevant. This formulation is then of the most
democratic form and removes the danger of being misled by the
different forms of space and time coordinates.  A difference between
those parameters arises in special situations such as when a Minkowski
background is re-introduced. This illustrates, again, that background
structures are to be eliminated as far as possible.

In addition to this extension from special to general relativity it is
believed that a further one is necessary to combine it with quantum
theory. A theory of quantum gravity in a reliable and completely
convincing form is not yet available, but from what we know it does
not seem to change much of the arguments presented here. It can even
eliminate potential problems such as that of singularities. At a
kinematical level, one can still imagine different possibilities
concerning the dimensionality\footnote{Background independent
formulations seem to agree on a lower dimensional kinematical nature
on microscopic scales
\cite{RS:Spinnet,ALMMT,SurfaceSum,LowDimDynTriag,FractalAsSafe}.} but
one still has the full parameter families corresponding to a
four-dimensional world when it comes to observables.

There are also conceptual advantages of a four-dimensional
understanding. If the world and its objects are four-dimensional, they
are simply there and do not need to become. There is then no need to
explain their origin, eliminating a difficult physical and
philosophical question.\footnote{Also for this question, quantum
gravity or cosmology seems to be affirmative: Initial conditions for
quantum cosmological solutions, which have traditionally been imposed
by intuitively motivated choices \cite{tunneling,nobound}, can arise
directly from the dynamical laws \cite{DynIn,Essay}. Thus, although
completely unique scenarios are difficult to construct,
four-dimensional dynamics can automatically select solutions and to
some degree eliminate additional physical input to formulate an
origin.}

\bibliographystyle{../preprint}
\bibliography{../Bib/QuantGra}

\end{document}